# Laser-driven ultrafast impedance spectroscopy for measuring complex ion hopping processes

Kim H. Pham[1], Scott K. Cushing[1*]

[1]Division of Chemistry and Chemical Engineering, California Institute of Technology, Pasadena, CA, USA.
*Correspondence to: scushing@caltech.edu

Superionic conductors, or solid-state ion-conductors surpassing 0.01 S/cm in conductivity, can enable more energy dense batteries, robust artificial ion pumps, and optimized fuel cells. However, tailoring superionic conductors require precise knowledge of ion migration mechanisms that are still not well understood, due to limitations set by available spectroscopic tools. Most spectroscopic techniques do not probe ion hopping at its inherent picosecond timescale, nor the many-body correlations between the migrating ions, lattice vibrational modes, and charge screening clouds—all of which are posited to greatly enhance ionic conduction. Here, we develop an ultrafast technique that measures the time-resolved change in impedance upon light excitation which triggers selective ion-coupled correlations. We also develop a cost-effective, non-time-resolved laser-driven impedance method that is more accessible for lab-scale adoption. We use both techniques to compare the relative changes in impedance of a solid-state $Li^+$ conductor $Li_{0.5}La_{0.5}TiO_3$ (LLTO) before and after UV to THz frequency excitations to elucidate the corresponding ion-many-body-interaction correlations. From our techniques, we determine that electronic screening and phonon-mode interactions dominate the ion migration pathway of LLTO. Although we only present one case study, our technique can extend to $O^{2-}$, $H^+$, or other charge carrier transport phenomena where ultrafast correlations control transport. Furthermore, the temporal relaxation of the measured impedance can distinguish ion transport effects caused by many-body correlations, optical heating, correlation, and memory behavior.

## I. Introduction

Advances in batteries, artificial ion pumps, and fuel cells require the departure from liquid-based ion conductors and toward solid-state alternatives, invoking a different understanding of fundamental ion transport mechanisms[1–3]. In perovskite-[4], LISICON-[5,6], and argyrodite-type[7,8] structures, ion transport is predicted to have complex correlated ion-ion[9], ion-phonon[10–14], and ion-charge screening cloud interactions[8,13,15–19] that occur on picosecond timescales. Such coupling phenomena are thought to be responsible for the fast ion transport observed in superionic conductors, with conductivities exceeding their liquid phase (>0.01 S/cm) and meeting the commercial metrics necessary to replace liquid electrolytes[20–23].

Techniques like Electrochemical Impedance Spectroscopy (EIS)[24,25], Nuclear Magnetic Resonance (NMR) techniques[26–28], and Quasi Elastic Neutron Scattering (QENS) techniques[29,30] are used to study ionic conduction across a range of timescales, with hopping frequencies typically inferred from linewidths[30,31] as shown in **Fig. 1**. EIS and Pulse-field gradient NMR probe time scales slower than site-to-site ion hopping and are thus more suitable for studying bulk and grain boundary hopping dynamics. NMR relaxometry experiments can probe ion-coupled dynamics on nanosecond timescales, but are more applicable to liquid conductors because the measured-coupling dynamics dominate longitudinal relaxation time measurements, rather than collective phonon-ion coupled vibrations found in the solid-state[28]. Neutron scattering techniques can probe picosecond resolution dynamics but provide time-averaged data rather than acquiring data in real-time[31]. Since EIS, NMR, and QENS do not probe local ion site-to-site hopping in real-time, the experimental data is often corroborated by Molecular Dynamics (MD) and other computational frameworks to predict superionic conduction mechanisms on picosecond timescales[31–33]. Thus, the limited temporal resolution accessed by current spectroscopy techniques also limits the ability to probe superionic transport mechanisms on its inherent timescales[31]. A spectroscopic technique that can probe hopping at ultrafast timescales up to picoseconds but also the complex ion coupling interactions would expand current knowledge in the field of solid-state ionics.

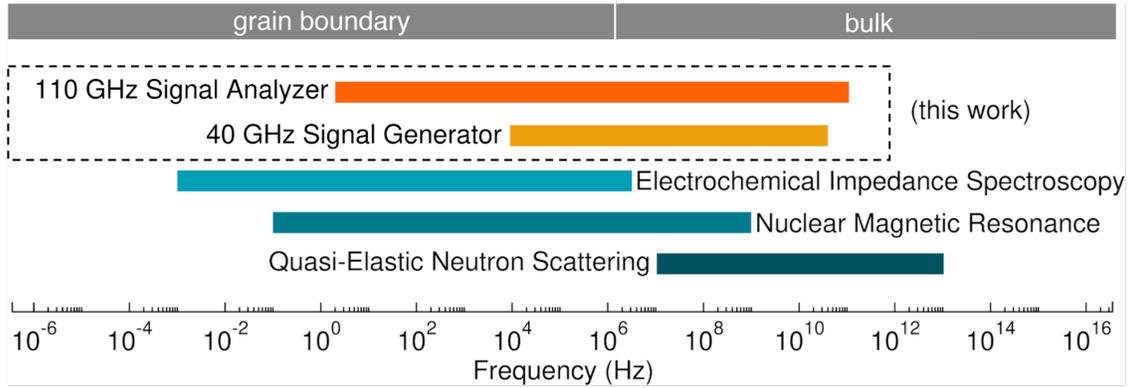

**Fig. 1.** Comparison of common techniques that probe ionic conductivity at different hopping time regimes, approximated by the experimental frequencies corresponding to each region in LLTO and the frequencies at which each technique operates. Although both Nuclear Magnetic Resonance (NMR) and Quasi-Electron Neutron Scattering (QENS) can infer site-to-site hopping regimes from linewidths, they do not directly measure ultrafast ion hopping and cross couplings like traditional ultrafast laser techniques. Our technique aims to probe hopping phenomenon in real time across the grain boundary, bulk, and site-to-site hopping regimes.

Ultrafast lasers can perturb phonon and electron interactions predicted to affect ion hopping[16]. Many-body type interactions influence ion conduction wherein the ion-host interaction can be perturbed using UV to THz light. For example, UV to visible light can photoexcite charge transfer transitions and create non-equilibrium carrier distributions to modulate screening effects[17–19,34] or even induce structural changes to enhance ion migration[35]. Near infrared to THz light can resonantly excite optical phonons[36], phonon modes[12,36–40], or trigger ion hopping[12]. Acoustic phonon modes can be selected by anharmonic or Raman interactions[12], or incoherently heated as a reference[37,38]. Several attempts in applying ultrafast measurements to the study of superionic conductors have been reported in literature[12,41]. In one study, Poletayev *et al* used impulsive near-resonant terahertz excitation to trigger ion hopping in single-crystal beta-alumina and monitored the transient birefringence as a Terahertz Kerr Effect (TKE) signal, which served promising despite the technique not directly probing the hop itself[12].

Even with the growth of new ultrafast spectroscopy techniques, ultrafast measurements are practically challenging because most ionic species do not have appreciable optical cross sections[16], nor do the devices that employ solid-state electrolytes operate under photoexcitation. Both challenges prevent the direct application of ultrafast pulsed laser techniques. While some successful cases exist, like 2D-Infrared spectroscopy of ionic liquids[42], a signal corresponding to bulk ion transport versus the relaxation of a photoexcited electronic transition, vibrational excitation, or displacement field, is difficult to discern and assign in solids because the many-body correlations and heterogeneity convolute the data[16]. Developing an accurate and direct probe for capturing picosecond ion motion dynamics remains challenging.

In this paper, we describe a laser-driven ultrafast impedance method that directly probes how many-body ion interactions, initiated by UV to THz light, directly change the ionic conduction behavior in a solid-state ion conductor at a resolution of 10 picoseconds. We accomplish these measurements by probing the S11 signal, or the reflected sample response signal upon impulse excitation while an applied field drives ion hopping further described in **Section II part A**. Steady-state signals were measured up to 110 GHz using a commercial Vector Network Analyzer while time-resolved measurements were conducted up to 32 GHz using a custom-built, version of a Vector Network Analyzer (VNA), composed of a signal generator and a real-time oscilloscope. In the time-resolved method, the signal generator applies a high frequency AC field to the ion conductor to drive ionic conduction. By applying high GHz frequency fields (40 to 110 GHz), the picosecond site-to-site hopping regime can be probed. The time resolved S11 reflectance signals are then measured as a function of sweeping the femtosecond pulsed from the UV to THz light. While collecting the S11 reflection measurements, the relative role of each many-body interaction with the hopping ion can be calculated, and comparing the respective relaxation times can provide insight into correlation and memory effects[12].

We apply the time-resolved, laser-driven, ultrafast impedance technique on a well-studied solid-state ion conductor $Li_{0.5}La_{0.5}TiO_3$ (LLTO) to investigate how a UV-band gap excitation affects the ionic conduction. The time-resolved ultrafast impedance measurements reveal that the band gap excitation causes an enhanced modulation spanning 100 ps, likely due to enhanced screening of the lattice cage. Recognizing the high cost of electronics that access picosecond time resolutions, we also describe and test a non-time resolved configuration that still probes the many-body-ion-couplings but at the sacrifice of time resolution. With a commercial impedance analyzer, the lower Hz to MHz frequencies can probe the grain boundary or bulk hopping regimes[24,25]. With the non-time resolved configuration, we observe that screening effects and phonon-coupled modes lead to enhanced ionic conduction. The non-time resolved configuration is compatible with both ultrafast lasers and continuous wave (CW) light sources and could therefore be more accessible to research groups across chemistry, material science, and biology. Both the time-resolved and non-time resolved techniques can be adapted for high throughput experiments to study coupling effects due to doping or elemental substitutions, paving the way toward better design principles for discovering new superionic conductors.

## II. Development of a time-resolved, laser-driven ultrafast impedance method

### A. General Concept

Ultrafast spectroscopy techniques typically rely on a pump-probe type experiment, where an ultrafast laser pump initiates a non-equilibrium population from which decay and scattering constants can be measured via the probe. In the linear limit, these timescales are assigned to a rare event near equilibrium, such as a thermally activated ion hops, by fluctuation-dissipation theory[43]. Although our technique superficially is akin to pump-probe spectroscopy with an ultrafast pump laser, our technique differs because the ultrafast laser pulse does not trigger the ion hop. Rather, a variable, high frequency, AC electric field drives mobile charge carrier hopping across a range of migration timescales in the material of interest. Applying an AC field (as opposed to DC) avoids polarization at the electrode interface, similar to how electrochemical impedance spectroscopy (EIS) operates[24,25]. Then, the ultrafast laser can drive electronic transitions to alter coulombic interactions[34], phonons to change the local energy minima along the hopping pathway[10], or adjacent ions to reveal possible migration pathways—all of which are predicted to couple to the hopping ion. Therefore, the experimental technique enables the study of ion-coupled migration phenomena at different hopping regimes at timescales set by the bandwidth of the AC field frequencies that can be generated as shown in **Fig. 1**.

In EIS, a sinusoidal, AC potential perturbs the sample of interest (**Eq. 1**) where V(t) is the potential as a function of time t, $V_0$ is the amplitude of the potential, and f is the frequency[44]:

$$V(t) = V_0 \sin(2\pi ft) \qquad \textbf{Eq. 1}$$

The interaction with the sample will cause a change in the phase and amplitude of the measured response (as current or potential) as shown in **Eq. 2**., where I(t) is the current as a function of time, $I_0$ is the amplitude of the current, and Θ is the phase angle. **Eq. 2** can also be rewritten for a change in potential.

$$I(t) = I_0 \sin(2\pi ft + \Theta) \qquad \textbf{Eq. 2}$$

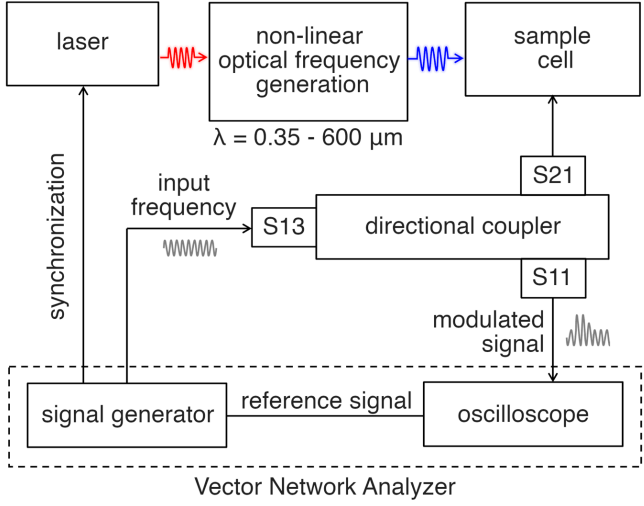

**Fig. 2.** Schematic of the time-resolved ultrafast impedance set up, including the laser, a representative nonlinear frequency generation process (to cover the UV to THz excitation range), VNA, and the sample cell.

The complex impedance can then be derived from the amplitude $Z = V_0/I_0$ and phase angle Θ, i.e., the phase shift between the measured $V(t)$ and $I(t)$ for a frequency $f$. While EIS cannot decouple conduction caused by multiple mobile charge carriers, EIS remains a useful tool to study ionic conduction in many solid-state conductors modeled with equivalent circuits that represent electrochemical processes with distinct time constants[25]. Experimentally, the Nyquist plot can be fit to an equivalent circuit and used to extract a corresponding resistance value associated with, for instance, ion hopping at grain boundaries and through a bulk grain. The resistance can be used to calculate the ionic conductivity with **Eq.3**:

$$\sigma = \frac{l}{R_{total} * A} \qquad \textbf{Eq. 3}$$

where $\sigma$ is the ionic conductivity, $R_{total}$ is the impedance that is treated as $Z_L$, l is the sample thickness, and A is the sample area.

In previous literature, FFT EIS[45–47], online EIS[48,49], and fast-time resolved techniques [47,50,51] have been developed to access faster hopping regimes. However, we report, to our knowledge, the first method that can reach up to 40-110 GHz due to recent advances in electronics, with previous reports only reaching 1-3 GHz[44,47,48,52,53]. Ultimately, the bandwidth of the signal generator and oscilloscope determines the temporal and frequency resolution, and using frequency extenders could soon reach THz frequencies.

To construct a time-resolved impedance measurement technique, a signal generator that can generate frequencies up to 40-110 GHz is used as the driving AC field to match the timescales of the ultrafast ion hopping in a bulk grain, shown in **Fig. 1**. The oscilloscope would then measure the potential change, current change, and phase shift for the same frequencies, all of which are used to derive impedance. Above tens of MHz, directly measuring the current or potential change can be on the order of sub picofarad and sub nanohenry[54], so the use of specialized electronics for high-speed detection of small currents and potential changes is critical. Vector Network Analyzers, for instance, can detect such small signal and provide measurements in the form of scattering parameters which derive from current and potential signals, measured in dBm[54]. Scattering parameters are useful measurements that help determine how a sample under test interacts with different waveforms by either transmitting or reflect some fraction of an input signal.

Specifically, the S11 scattering parameter measures the ratio of the change in the signal input or power reflected from the port 1 input ($b_1$) and the power or signal incident to the port 1 input ($a_1$), assuming that the boundary condition $a_2 = 0$ (port 2 output) holds as shown in **Eq. 4**. The S11 parameter can also be represented as the reflection scattering parameter Γ. The S21 scattering parameter measures the ratio of the signal transmitted to the load and the signal transmitted from the load as shown in **Eq. 5**.

$$S_{11} = \left.\frac{b_1}{a_1}\right|_{a2=0} \qquad \textbf{Eq. 4}$$

$$S_{21} = \left.\frac{b_2}{a_1}\right|_{a2=0} \qquad \textbf{Eq. 5}$$

The measurement of the S11 to obtain part of the sample impedance can be described by **Eq. 6-**

**8**[54,55], where VSWR is the voltage standing wave ratio, or the efficiency of transmission of a radio frequency power source to the load through a transmission line, S11 is the reflection scattering parameter, Γ is the reflection coefficient, $Z_0$ is the reference ohm (typically 50 ohm), and $Z_L$ is the sample impedance or load.

$$\text{VSWR} = \frac{1 + |S11|}{1 - |S11|} \quad \textbf{Eq. 6}$$

$$\Gamma = \frac{\text{VSWR} - 1}{\text{VSWR} + 1} \quad \textbf{Eq. 7}$$

$$Z_L = Z_0 * \frac{1 + \Gamma}{1 - \Gamma} \quad \textbf{Eq. 8}$$

In an experimental set up as shown in **Fig. 2**, an input signal at port S13 can be sent to the sample at port S21 via a directional coupler. The reflected signal from the sample can be measured at S11 to determine how the sample interacts with the input frequency originally at S13. Since the S11 is measured directly from the VNA via a directional coupler, the $Z_L$ can be calculated and subsequently the ionic conductivity. The S21 parameter can also be calculated to understand how the sample in study changes the input waveform[55]. From the S11 and S21 scattering parameters, the sample impedance can be measured to determine the complex conductivity.

To connect the sample to the directional coupler, a vertical launch is used which contains a conductive pin that electronically contacts the sample in series in a transmission line geometry[56] as shown in **Fig.3**. The experimental geometry can be modeled as a simple circuit like EIS, where the sample has an associated resistance and capacitance in parallel with the resistance of the short as described in previous literature and mentioned in more depth in **Section III part D**[56]. Additionally, by adding the copper plate as a short in the circuit, the conditions required for the S11 measurement is met or i.e. ensuring that $a_2 = 0$. A microstrip line type geometry is configured for measurements below 32 MHz measurements in this work, but can also be optimized for measurements above 1 GHz. In a transmission line or microstrip geometry, the S21 can be measured to obtain the transmission coefficient which can be used to determine the complete complex impedance together with the measured S11. In this work, we focus on the experimental set up of only S11 measurements which is described in more detail in the next section.

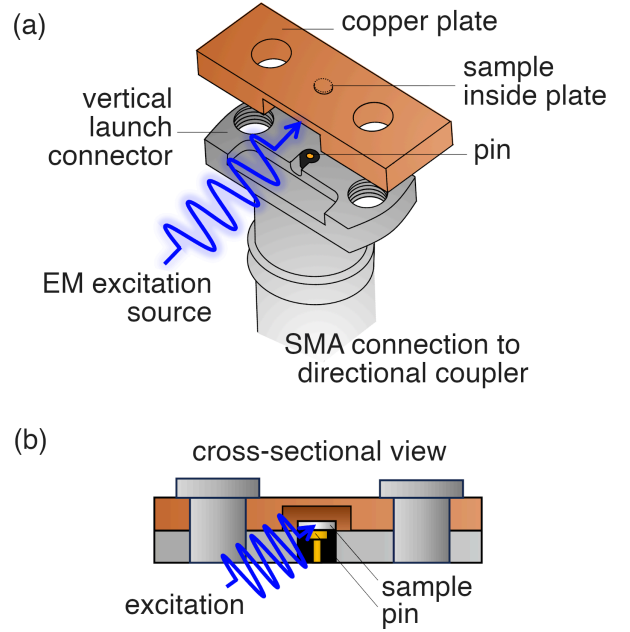

**Fig. 3. (a)** Schematic of vertical launch with an exposed pin that makes electrical contact with the sample under study. **(b)** A cross-sectional view of the sample holder geometry. A cavity is machined out from the side of the copper plate, exposing the sample that sits inside a pocket 300 microns in depth and 2 mm in diameter. The cavity allows the sample to be irradiated by an excitation source.

### B. Instrument Construction

A schematic of the instrument is shown in **Fig. 2**. For the excitation laser source, a Ti:Sapphire laser oscillator and amplifier (Legend Elite and Astrella-V-USP) from Coherent are used to create the visible to THz light from non-linear optical frequency generation methods. UV to NIR light is created using an optical parametric amplifier (TOPAS) and NirUVis from Light Conversion. 350 nm light is generated using fourth harmonic generation from the TOPAS and NirUVIS. A 349 nm laser diode (Explorer One 349-60, Spectra-physics, 60 uJ pulse energy, 1 kHz repetition rate) is also used for comparison with UV light. Difference frequency generation is used to cover the 5 to 15 μm range. THz light is generated using a DAST crystal and the 1400 nm output of the optical parametric amplifier.

Experimentally, 1-25 mW of average power is used for the 1 kHz laser and 500 mW of average power for a CW laser depending on the sample and focused into a beam of 200 – 300 microns in diameter for the EIS measurements. For the time resolved measurements, the 800 nm pulsed beam was focused from a beam approximately 3,000 microns in diameter through a 7.56 cm focal lens and a 350 nm pulse beam approximately 800-900 microns in diameter.

A 2 Hz – 110 GHz Keysight N9041B UXA Signal Analyzer was employed as a VNA for steady-state measurements. To measure the time-resolved impedance signal, a 40 GHz, Keysight, N5173B EXG X-Series Microwave Analog Signal Generator coupled with a Keysight UXR0334A Infiniium UXR Real-time 33 GHz, 128 GSa/s, 13.3 ps rise-time Oscilloscope is used. The noise floor determines the lowest differential signal that can be measured and depends on the exact signal generator and oscilloscope used. With the described equipment, a signal to noise ratio of $10^{-4}$ -$10^{-6}$ was possible. For the non-time resolved method, a 1260A Solartron impedance analyzer is used to measure the < 32 MHz frequency ranges and conduct EIS.

For S11 reflection measurements, the VNA generates an AC signal that transmits to the sample, or load, via a directional coupler with an SMA connection into the sample and a copper short at port S21 with circuit modeling reported from previous literature[53,57,58] as shown in **Fig. 2**. A cavity is drilled out from the side of the copper short to allow laser excitation during the time-resolved measurements as shown in **Fig. 3a**. The powder sample is densified into a 2 mm diameter pellet under high force (up to 2 tons), annealed to achieve at least 80% of its theoretical density (specific to the composition), and subsequently sanded to fit inside the 300 μm well. In addition to minimizing air gaps, the sample must make physical contact with the pin inside of the vertical launch connector to create a resonator, which is critical for accurate measurements, as shown in **Fig. 3a and 3b**. The RF input is sent to the sample at the S21 port from the S13 port and the resulting perturbation is measured at the S11 port which measures the reflected wave. To physically measure the S11 parameter, the sample adopts a transmission line configuration with the internal pin of the vertical launch, enabling high frequency transmission to over 10 GHz[54]. The electrical connection between the sample, contacting pin, and the oscilloscope is established with a co-axial cable with appropriate adapters rated for GHz frequencies, such as SMA connections.

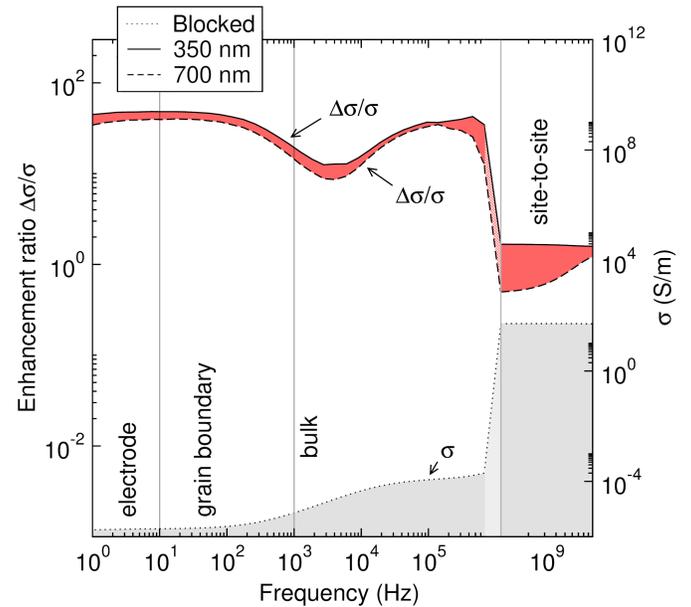

**Fig. 4.** The enhancement ratio or change in ionic conductivity of LLTO from 2 Hz to 110 GHz. Excitation of a charge-transfer transition reduces the coulombic hopping barrier, increasing conductivity (black solid line). The change, shown as the red area, is greater than optically heating the lattice (black dashed line). The dotted line shows the conductivity of LLTO with the blocked laser. The light red and gray area indicates unmeasured regions in the data.

The laser clock can be used as an external reference to phase sync with the custom VNA, but in some cases, syncing can be difficult because of excessive clock jitter from the pulse laser arrival and signal generator. Many signal generators only accept a 10 MHz reference signal that a laser clock will not output. Even after frequency dividing the laser clock output to achieve 10 MHz, the resulting signal may not have sufficient phase stability. An IQ demodulation approach could help to then extract the amplitude and phase information. Here, a 35-picosecond photodiode is used to phase sync the signal by triggering off the 1 kHz laser. Next, the S11 measurement is conducted for a combination of varying laser excitation frequencies and signal generator frequencies, depending on the experiment.

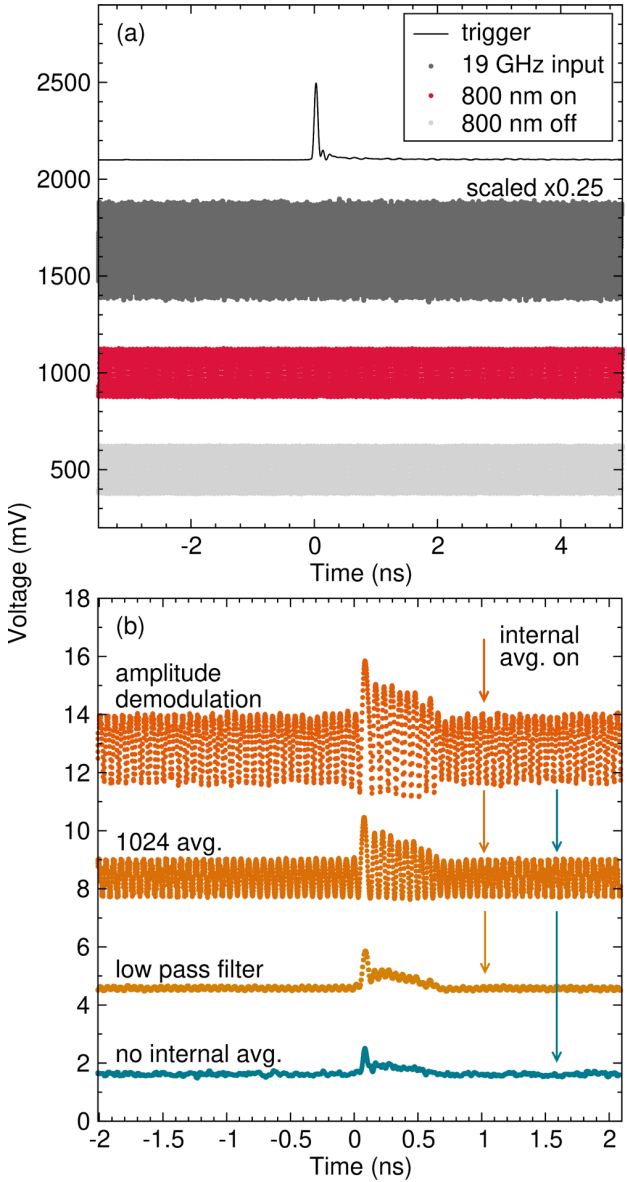

**Fig. 5. (a)** A voltage versus time trace of the 800 nm used as the trigger source (black), signal from the 19 GHz 19 dBm signal at the S21 port (dark gray), signal from sample upon 800 nm excitation (red), and blocked sample at the S11 port (light gray). The signals show the accumulated waveform over the collection time of the real-time oscilloscope. The sine wave is not phase synchronized with the laser because of the laser oscillators timing jitter being too high. **(b)** The amplitude demodulated signals from the raw red trace obtained in (a), showing the sequential application of functions to recover the low frequency demodulated signal and optimize noise. Note that the amplitude demodulation, 1024 avg., and low pass filter functions all have the internal hardware averages set to 1024 avg. while the bottom teal trace has no internal hardware averages on but contains the same applied functions showed in the above three traces.

### C. Application of steady state measurements UV to visible excitation with the steady-state VNA

The 2 Hz – 110 GHz Keysight N9041B UXA Signal Analyzer is used to perform steady state measurements on LLTO upon CW 350 nm and 700 nm excitation to explore the conceptual capabilities of the instrument before the application of time-resolved experiments. The VNA is, at its essence, a coupled signal generator and oscilloscope, enabling direct comparisons between steady-state and time-resolved measurements. The S11 signal is Fourier-filtered to remove the carrier signal frequencies and is used to calculate $Z_L$ using **Eq. 6 – 8.** The $Z_L$ can then be used to calculate the ionic conductivity using **Eq. 3**.

LLTO is used as the test sample for the developed instrument due to its stability in air[59] and well characterized ion migration pathway[60], band gap energy[61], vibrational modes[62], and phonon modes[38,63], all of which are predicted to be involved in ionic conduction. $Li^+$ conduction in LLTO is mediated by adjacent vacancies in between bottlenecks formed by four oxygens from four corner shared $TiO_6$ octahedra[60,62]. Screening effects have been predicted to aid ionic conduction by minimizing electrostatic interactions between the host lattice and migrating ion, enabling fast ion migration for several solid-state $Li^+$ and $O_2^{2-}$ conductors[17–19,34]. In LLTO, a ligand-to-metal charge transfer transition can occur upon the 2.1 eV band gap excitation which promote electronic carriers from the O 2p orbitals to the Ti 3d orbitals[61]. By transferring charge density from the O 2p orbitals to the Ti 3d orbitals, the electrostatic hindrance in the ion conduction pathway is likely reduced, giving insight into the ion-electron interactions. The effects caused by the UV are compared to the 700 nm light which is sufficiently above the band gap energy and should only induce thermal effects as reported in previous work[38].

LLTO is synthesized according to literature using a solid-state synthesis preparation with $La_2O_3$, $Li_2CO_3$, and $TiO_2$ as precursors. The sample was calcined at 800°C for 4 hours and heated to 1200°C for 12 hours at a 1˚C ramp rate[59]. The LLTO was structurally and electrochemically characterized using powder x-ray diffraction and EIS as shown in previous work[38]. To test the role of screening on ionic conduction, the bandgap of LLTO is optically excited through the cavity accessing the LLTO sample in the vertical launch geometry shown in **Fig. 3**. Although the application of the vertical launch geometry was suitable enough for proof-of-concept measurements, noise above 10 GHz was observed, potentially due to contact issues and pellet density. Further optimization of the cell design would enhance the signal to noise ratio in future studies.

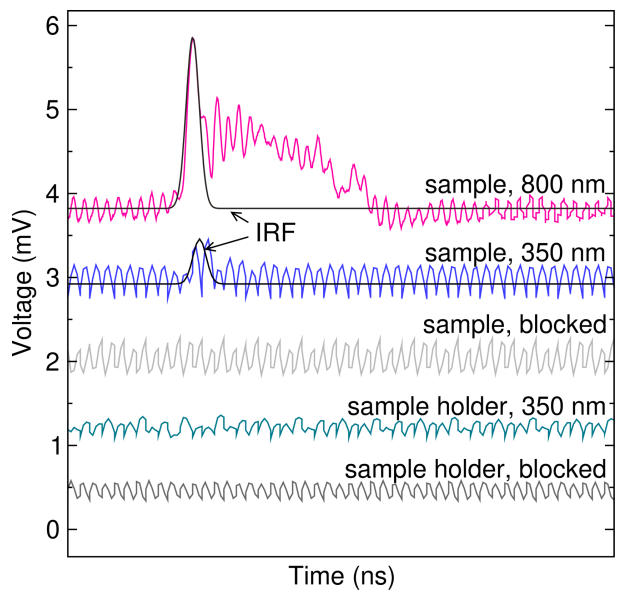

**Fig. 6.** Comparison of signal responses from 800 and 350 nm excitations with the sample holder with and without illumination, in addition to the sample without illumination. The black lines represent the instrument response function, with a bandwidth cutoff of 33 GHz. The input carrier frequency is at 28 GHz and 19 dBm for the 800 nm measurement and 10 dBm for the 350 nm measurement. The amplitude was optimized to obtain the best signal to noise ratio. The 800 nm excitation was offset by approximately -2.3 ns to compare the characteristics signal peaks with the 350 nm.

The steady-state response was measured and used to calculate and plot $\Delta\sigma/\sigma$ as a function of frequency as shown in **Fig. 4**. which combines the low frequency range of an impedance analyzer and higher frequency ranges of the Vector Network Analyzer, the plot shows the enhancement in conductivity of LLTO after photoexcitation with a 350 nm CW laser at approximately 500 mW. The shaded red region in **Fig. 4** shows the difference in the changed ionic conductivity due to laser heating of the incoherent phonon bath with 700 nm light at similar powers versus the modulated $Li^+$-electron coupling from the 350 nm bandgap excitation. The largest increase in the enhancement ratio is observed in the site-to-site hopping region and is likely due to the shift in charge density from the O 2p to Ti 3d orbitals which we predict to lower the activation energy for the ion hop. The differences observed due to the 350 nm and 700 nm at the grain boundary and across the electrode-electrolyte surface regions is also observed and are likely related to the population of thermal baths.

Although the generation of electronic carriers can convolute the final $Li^+$ mobility values typically determined through impedance methods like EIS, many reports suggest that enhancements in ion migration are not exclusively due to the newly generated electronic carriers[17–19]. Thus, we believe our observations of enhanced ion migration due to this charge transfer process holds, after accounting for heat generation and electronic carrier generation. Even if electronic conduction effects are also present, the experiments described still prove the general concept of the instrument in its steady-state implementation.

### D. Application of time-resolved measurements with custom VNA

The custom, time-resolved version of the VNA comprising the 40 GHz signal generator and 33 GHz real-time oscilloscope is used to measure the time resolved enhancement in ionic conduction in LLTO due to band gap excitation. The sample cell configuration shown in **Fig. 3** is used to conduct the measurement and follows the experimental configuration shown in **Fig. 2**.

**Fig. 5a** shows the direct AC voltage measurement at S11 with and without the 800 nm with a 19 GHz, 19 dBm signal applied to the sample. The 800 nm excitation uses 20 mW of average power focused with a 76 mm focal lens to approximately 50 microns in diameter. A very small modulation is barely detectable on the AC voltage measurement occurring between 0 – 200 ps without internal averaging, but ultimately results in an amplitude modulated signal where the input RF frequency from the signal generator acts as the high frequency signal carrier and the modulation is low frequency in nature. To improve the noise, the internal hardware averaging is applied and set to 1024 averages to improve the signal quality. To extract the low frequency modulation from the high frequency carrier signal and achieve high signal to noise measurements, an amplitude demodulation function is applied afterward such that an envelope detector removes the carrier signal and leave behind the low frequency modulation i.e. the signal caused by the light excitation[64]. As shown in **Fig. 5b**, the amplitude demodulated signal reveals an impulse peak followed by a growth and decay of the signal over 1 ns. The noise can be further reduced by subsequently averaging the amplitude demodulated signal with 1024 averages. The low-pass filter can be applied on top of the averaged signal to further reduce the noise but should be used with caution to avoid completely removing a signal that would otherwise be present. To our knowledge, the demonstration of the ultrafast impedance-type technique up to high GHz frequencies revealing a thermal decay lasting a nanosecond at tens of picosecond time resolution is one of the first of its kind. To confirm that the 350 nm excitation interacts with LLTO differently than the 800 nm excitation based on the proposed charge transfer excitation mechanism, the 350 nm beam was also excited onto the sample with 13 mW of average power onto a spot size of approximately 800-900 microns in diameter. Due to the low efficiency of fourth harmonic generation, a similar average power of 20 mW could not be achieved like the 800 nm source and necessitated a larger spot size to excite more of the sample volume.

As shown in **Fig.6**, the impulse of the 350 nm results in the appearance of an impulse limited response, similar the 800 nm. However, a distinct decay with a time constant of ~ 100 ps is the only feature for the 350 nm signal. In the 800 nm case, there is a growth and decay after the impulse response. The 800 nm response could be characteristic of the rise and decay of the acoustic phonon bath (generally >100 ps) following the optical phonon bath decay (generally <100 ps). The 350 nm excitation may support this, as the charge-transfer related modes interact more strongly with optical phonons as compared to acoustic phonons. Even though our observations and hypotheses are preliminary and require more studies to conclude the true nature of the temporal features, we are still able to demonstrate the ability to measure picosecond ion hopping dynamics.

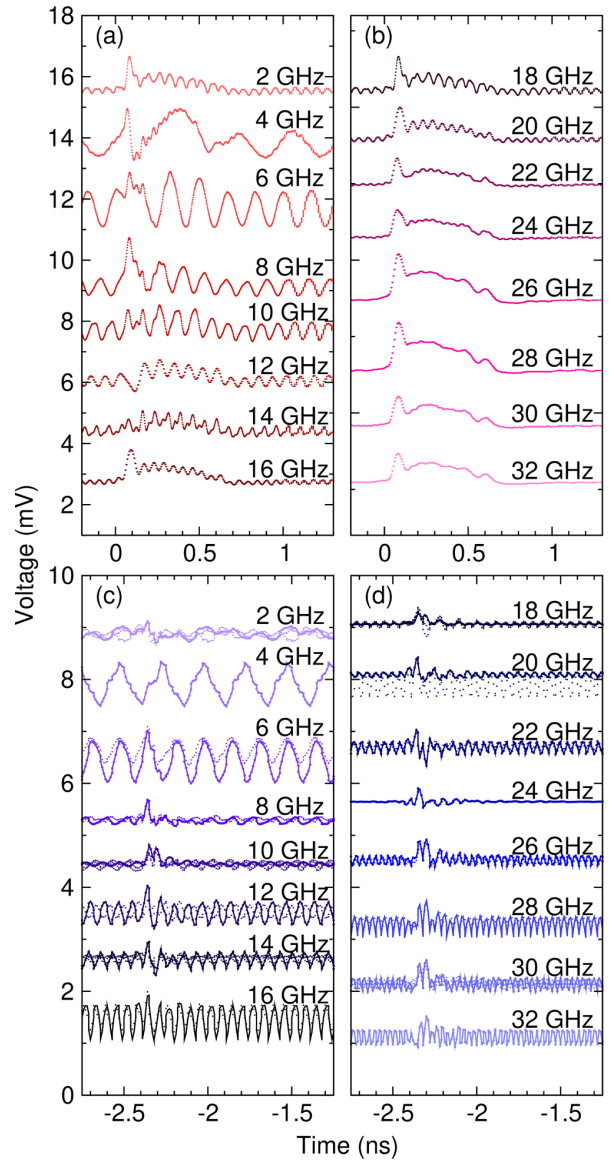

**Fig. 7. (a)** The sample signal under 800 nm excitation between 2-16 GHz input frequencies at 19 dBm and **(b)** between 18-32 GHz at 19 dBm. **(c)** The sample signal under 350 nm excitation between 2-16 GHz input frequencies at 10 dBm and **(d)** between 18-32 GHz at 10 dBm.

To obtain measurements like the steady state measurements described in **Section II part C**, a frequency sweep can be initiated on the custom VNA to show how the sample signal changes with carrier frequency. As a proof of concept, we conduct a frequency sweep from 2 - 32 GHz for both the 800 nm excitation and the 350 nm excitation as shown in **Fig. 7a-b** and **Fig. 7c-d**. For the 800 nm excitation frequency sweep data, the characteristic sample feature is shown to persist between 16 to 32 GHz while at frequencies below 16 GHz, the characteristic shape disappears. A low pass filter is applied to the measurements for the **Fig. 7a** and **7b** to optimize the signal to noise. The bandwidth is adjusted so that the value is less than 50-70% of the carrier frequency signal which helped reduce noise while not removing any features associated with the sample signal. For the frequency sweep of the sample excited at 350 nm, the characteristic signal seems consistent between 26 – 32 GHz and variable in shape below 26 GHz. We hypothesize that the characteristic signal becomes more distorted at lower frequencies because of the sample holder geometry's ability to amplify the resonant signal in addition to our current setup using excessively lossy cables for lower frequency ranges. Further work is needed to explore and optimize the described experimental methodology to extend measurements to lower frequencies in addition to measurements at high GHz frequencies through optimization of the sample holder and sample size.

Ultimately, the time resolved data at the picosecond timescale gives insight to how site-to-site hopping is influenced by screening at the local scale which has, to our knowledge, never been demonstrated previously, and would expand existing knowledge on how the unique local structure of hopping channels collectively influence ionic conduction.

### III. Non-time resolved laser driven impedance: Application with commercial impedance analyzer

#### A. Practical differences between non-time resolved and time-resolved methods

The hundred GHz bandwidth accessed by a time-resolved VNA can be cost prohibitive due to the advanced electronics, so we also test a cost-effective, non-time resolved, laser-driven impedance set up. A CW light source is coupled to a 1260A Solartron impedance analyzer to measure changes in the complex impedance between 1 Hz to 32 MHz, encompassing grain boundary and the tail of the bulk conduction regimes in many materials. While a CW light across the UV to THz is ideal in this case, generation of light with sufficient power densities to cover the range of electronic to vibrational modes is challenging. The more commonly used Ti:Sapphire laser and its nonlinear frequency mixing can easily access UV to THz frequencies with high power, however, the 1 kHz repetition rate combined with the slower EIS measurements means that any measured dynamics are time-averaged and a thermalized value must be obtained, lowering sensitivity. The incident power, power density, and penetration depth of each wavelength must be carefully considered now. Rigorous treatment of the control data is necessary because of duty cycle averaged impedance data when using a pulsed laser. We found that an optimal solution exists by normalizing the change in impedance per change in heat using our custom heating cell shown in **Fig. 8**. which we describe in **Section III part B**.

#### B. Normalization treatment of non-time resolved laser-driven impedance data

The normalized impedance per change in heat accounts for the total absorbed power independent of the strength of a transition and is applied specifically for our demonstrated measurements with EIS because the resulting data acquired is duty-cycle averaged over milliseconds of time. In the time-resolved case, the 10s of picosecond resolution of the ultrafast impedance measurement can distinguish heating effects from other optical effects and thus do not require the described normalization treatment. Since the complex impedance measurements are taken at equilibrium after several minutes, the penetration depth relative to the surface electrode is factored into the final value. Therefore, the normalization methodology allows for direct comparison between the laser heating and DC heating to accurately calibrate the data baseline.

We accomplish the normalization by first constructing a calibration curve with the custom heating cell shown in **Fig. 8** using a TC-48-20 OEM temperature controller and thermocouple. The current and voltage that is measured across the cell and through the sample is used to calculate the power input into the sample based on Watt's law P = IV. By plotting the calculated P versus the temperature read out from the TC-48-20 OEM software, a calibration curve can be built to determine a temperature associated with a power input into LLTO as reported in previous work[38]. The normalization approach was found to give similar percent changes as normalizing by power density, proving the validity of the proposed method. Despite the experimental limitations between using the CW and pulsed laser excitations to accurately interpret the data, non-time resolved measurements with the commercial impedance analyzer can be accomplished and still provide valuable information on coupling dynamics in a variety of solid-state materials without needing custom electronics.

## C. Cell Design

For the non-time resolved measurements up to 32 MHz frequencies, the custom cell was constructed with a ceramic heating plate with temperature-controlled operation using the corresponding software from the TC-48-20 OEM heater and thermistor as shown in **Fig. 8.** The heating cell is placed inside a faraday cage made of copper mesh with a 1.4 mm wire spacing for all experiments to reduce noise from electromagnetic interference[24]. A THz transparent window can be designed with a variety of organic materials and crystalline materials[65]. For the other visible/UV light sources, a quartz window that is transparent between 190 nm – 2500 nm is used. A windowless set up can alternatively be employed for non-air sensitive materials. The sample itself is annealed and densified as described previously, and a blocking electrode, such as Au, Ag, Pt, or Pd, is sputtered onto the pellet with a mask to create a gap-electrode geometry. Sputtering Au electrodes on the same plane of the sample pellet with a gap geometry shown in **Fig. 8**, rather than fully sputtering both planes, minimizes the volume of sample that is probed during the impedance measurement, rendering the effect of differences caused by optical penetration depth and spot size to be more negligible. Thin film samples can also be used so that optical penetration depths are negligible. The developed in plane electrode configuration is compatible with the time resolved set up and can also test grain boundary and bulk conduction effects, with grain boundary contributions dominating the overall ionic conductivity when used as solid electrolytes in all solid-state batteries.[34]

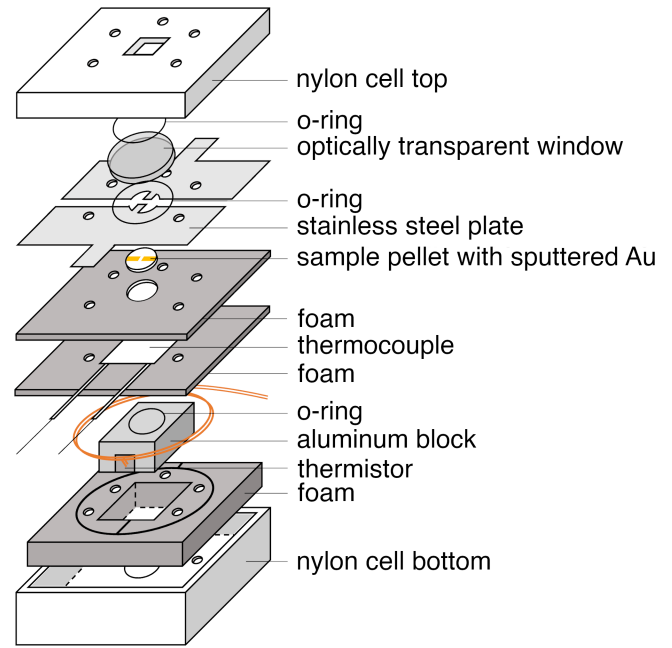

**Fig. 8.** Electrochemical heating cell set-up to obtain the power-to-temperature calibration curve and collect EIS data below 32 MHz with the 1260A Solartron. The cell components are compressed and held together with screws that fit through the five holes depicted. Each screw is secured with wingnuts.

## D. Application of non-time resolved measurements with impedance analyzer

The change in impedance caused by UV-excitation described previously in the time resolved case is demonstrated again here with the non-time resolved methodology, shown in **Fig. 9**. The data in **Fig. 9** is reported as a change in impedance rather than ion conductivity because of the change in sample geometry (in-plane pellet surface versus through pellet, or **Fig. 8** vs. **Fig. 3**, respectively) which prevents accurate determination of the surface thickness and area required to calculate ionic conductivity. However, since all pellets were prepared with the same thickness and diameter, the relative changes in impedance due to excitation are reproducible and reliable within error over a set of three pellets and three trials per pellet. The equivalent circuit fit for the bulk conduction feature (**Fig. 9a**) and grain boundary feature (**Fig. 9b**) is shown along with the corresponding Nyquist plots, demonstrating good agreement. Specifically, the grain boundary feature is fit to an R1+ R$_{GB}$/Q2 circuit while the bulk feature is fit to an R1+ R$_{Bulk}$/Q2. The R values corresponding to the grain boundary (R$_{GB}$, ranging from 1200 – 900 kΩ) and bulk hopping regimes (R$_{Bulk}$, ranging from 18 - 13 kΩ) are assigned based on capacitance values obtained from the circuit fit. The final R$_{GB}$ and R$_{Bulk}$ extracted from the fit is plotted against UV power in **Fig.9 c-e**, demonstrating that the measured change in impedance is linearly proportional to power. The sequential shift towards decreased impedance upon increasing UV light average power indicates that screening effects promoting facile ion conduction, mirroring the results described previously in **Section II**.

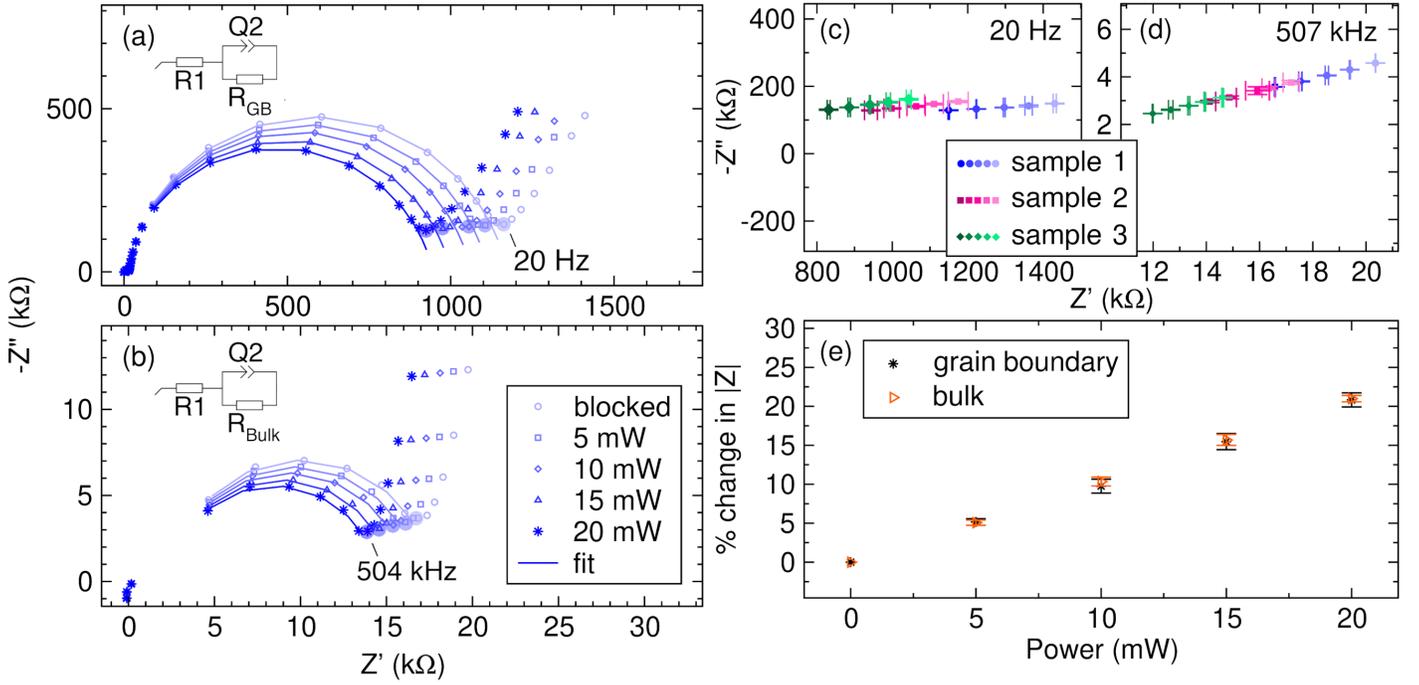

**Fig. 9.** Change in impedance upon band-gap excitation with 349 nm light between 5 – 20 mW. **(a)** Grain boundary semi-circle fit to a R1+R$_{GB}$/Q2 circuit. **(b)** Bulk impedance semi-circle fit to a R1+R$_{Bulk}$/Q2 circuit. **(c)** Grain boundary and **(d)** bulk replicates across three samples showing the change in impedance upon 349 nm light excitation over 5 – 20 mW at a frequency corresponding to the respective intercept of the semi-circle feature. **(e)** Percent change in impedance as a function of 349 nm laser power shows a linear response for both the grain boundary and bulk.

**Fig. 10(a)** shows the enhancement ratio ΔR$_{bulk}$/ΔK of LLTO due to heating effects and is compared to the enhancement caused by a range of excitation frequencies including 800 nm light, mid-IR light, and THz frequencies measured in previous work[38] as well as the 350 nm light excitation from this work shown in **Fig. 10(b)**. Briefly, the chosen excitation frequencies are used to determine changes in the enhancement ratio due to optical heating, resonant La-O bond excitation, and phonon mode excitation respectively. *Ab initio* calculations were used to determine the theoretical contribution of different THz modes up to 25 THz to the Li$^+$ hopping trajectory. The computational results reveal that the majority of the modes that are excited experimentally (up to 6 THz) are primarily rocking octahedra modes which provide over 40% of the energy required to initiate a Li$^+$ hop in LLTO[38].

The relevant absorption features from the UV to THz are shown in **Fig. 10(b)** as the magenta lines.

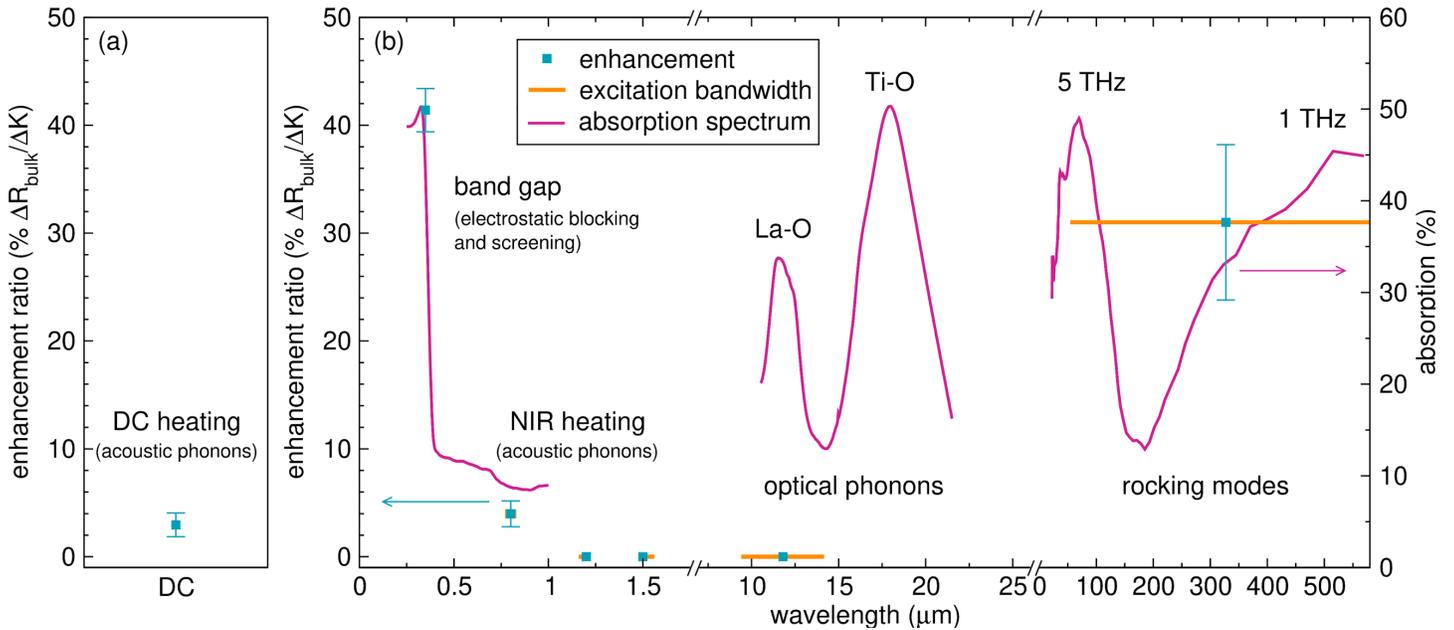

**Fig. 10.** Enhancement in ion migration due to thermal contributions and illumination across the near-IR (NIR), mid-IR (MIR), and THz light. The data at 0.35 μm was obtained from this work while all other data was obtained from an earlier study. **(a)** The enhancement in migration from DC heating corresponds to incoherent heating of the acoustic phonon bath. The change in $R_{bulk}$ per change in sample temperature (K) is represented by the cyan point. **(b)** Enhancement caused by light excitations at frequencies from 0.350 μm to 600 μm. The purple lines represent the absorption spectra of LLTO from NIR to THz. The width of the horizontal orange bar represents the spectral width of the excitation pulse. The NIR enhancement corresponds to incoherent heating of the acoustic phonon bath. The MIR excitation corresponds to coherent driving of optical phonon modes, which was inconclusive for the same excitation density as the THz modes. The THz light coherently drives highly contributing modes, showing large relative enhancement.

The cyan points represent the enhancement ratio of the bulk impedance after laser driving within the corresponding frequency range, while the width of the orange bar represents the bandwidth of the excitation source. The DC and NIR heating plots serve as the control since both excitation sources incoherently heat the material through the acoustic phonon bath. Like described previously for **Fig. 4,** the UV excitation likely changes the electrostatic blocking in the lattice cage by a ligand-to-metal charge transfer from the O 2p valence band orbitals to the Ti 3d conduction band orbitals, causing the ten-fold enhancement. The UV excitation confirms that a significant electrostatic hindrance to the ion hopping through the lattice cage exists as discussed in other ion conductors like LGPS[6], and can be alleviated with a band-gap excitation. The result is consistent between both the time-resolved and non-time resolved methods, further validating the accuracy of the techniques Although the enhancement ratio also likely has some contributions from electronic carriers generated by the UV light, we believe the measured signal also majorly constitutes ion migration. The experimental determination of the exact contributions of electron versus ion migration is currently in progress.

In the mid-IR range, a signal is not detected for photoexciting the La-O and Ti-O optical phonon branches with similar powers as the THz rocking modes, but a higher power DFG unit would provide more definitive evidence. Finally, the enhancement shown in the THz region represents the broad excitation of numerous $TiO_6$ rocking modes within 0 – 6 THz [38,63] and proves to be a dominant term in changing the bulk ion hopping in addition to screening. The strong coupling interactions between the $TiO_6$ octahedral rocking modes and the migrating $Li^+$ are responsible for the enhanced ion migration observed experimentally.

**Fig. 10** demonstrates that the relative contribution of the many-body interactions to ion migration compared to incoherent heating shown in **Fig. 10(a)** can be measured and compared when normalized by heat. For the test data provided here, in the case of LLTO, **Fig. 10(b)** shows that the THz rocking mode is a dominant vibrational mode as compared to optical phonons or the rest of the acoustic phonon bath as shown experimentally and supported by *ab initio* calculations. Although the CW excitation source would allow measurements of absolute changes in ionic conduction with respect to excitation frequency, the pulsed-average method still provides valuable information on the relative role of couplings on ionic conduction. Even with the benefits of a CW source to achieve true equilibrium, reaching the far-IR to THz range with sufficient power is difficult.

To further minimize additional costs, lamps and incoherent light sources could be adopted for a more straightforward approach without the need for specialized optics knowledge or the cost of an ultrafast laser, given the sample is normalized by heat as demonstrated in **Fig. 10**. Photo-modulated spectroscopy has been adopted for many applications including fast charging in batteries[35] and modulating ion hopping[34,41], and we believe that similarly, our measurements can be adopted to investigate charge transport systems.

In sum, the differences in the enhancement ratio values between the hopping time regimes across a broad range of frequencies shows how each regime can behave differently and warrants the exploration of the unique couplings that have been predicted to influence ion migration. By comparing the relative enhancement ratios at different hopping regimes due to a series of targeted ion-phonon, ion-electron, and ion-ion excitations, one could reveal couplings that dominate the ion conduction mechanism and pave the way for targeted design of superionic conductors.

## IV. Conclusions

The time-resolved, laser-driven ultrafast impedance technique presented here can directly measure ion hopping on picosecond and longer timescales while comparing the absolute and relative role of ion couplings to phonons, electrons, and other ions. Our proposed technique overcomes the challenges of other ultrafast time resolved approaches by utilizing the laser as a probe in an AC measurement, rather than using the laser to initiate ion conduction as a pump source. The described method ensures that the resulting transient or signal directly probes ion conduction. Although this study focuses on one type of $Li^+$ conductor with low contributions to electronic conductivity, extensions to mixed ion-electron conducting systems is certainly possible. Additionally, the cost-effective, photo-modulated or action spectrum-like method provides a more lab-accessible route to probe complex ion-couplings, which can leverage the use of cost-effective light sources like a high-power, broad-spectrum lamps with a monochromator. Even though time-domain information about couplings and correlations are lost, the relative impact of different electronic and vibrational interactions can still be compared.

## Acknowledgements

Financial support was provided by the National Science Foundation and Air Force Office of Science & Research (FA9550-21-1-0022). We thank David Weldon and Luis J. Hernandez from Keysight for their assistance with the signal analyzer, signal generator, and oscilloscope. We thank Ricardo Zarazua and Martin Mendez at the Chemistry and Chemical Engineering Machine Shop for machining the custom heating cell in this work and the copper sample holder respectively. We thank Amy Lin for her assistance in preparing LLTO pellets for the time-resolved measurements.